\documentclass[a4paper,fleqn]{cas-dc}

\usepackage[style=ieee,minbibnames=3,natbib=true]{biblatex}
\usepackage{color,soul}

\def\tsc#1{\csdef{#1}{\textsc{\lowercase{#1}}\xspace}}
\tsc{WGM}
\tsc{QE}

\begin{document}
\let\WriteBookmarks\relax
\def\floatpagepagefraction{1}
\def\textpagefraction{.001}

\shorttitle{}    

\shortauthors{}  

\title [mode = title]{Efficient cosmic ray generator for particle detector simulations}  




%

\author[1]{David Díez-Ibáñez}[orcid=0000-0002-0797-1876]

\cormark[1]


\ead{daviddiez@unizar.es}

\ead[url]{https://github.com/DavidDiezIb}

\credit{Conceptualization of this study, Methodology, Software}

\affiliation[1]{organization={Center for Astroparticles and High Energy Physics (CAPA), Universidad de Zaragoza},
            city={Zaragoza},
            country={Spain}}
            
\author[1]{Luis Obis}[orcid=0000-0002-7990-2060]

\cormark[1]


\ead{lobis@unizar.es}

\ead[url]{https://github.com/lobis}

\credit{Conceptualization of this study, Methodology, Software}


\cortext[1]{Corresponding authors}



\begin{abstract}
    Traditional cosmic ray simulations commonly employ the Monte Carlo method to randomize the energy and direction of each simulated particle, often employing simplified or uncorrelated distributions. The flux of cosmic rays is modelled as incident particles originating from a plane above the object of interest (e.g., detectors in particle physics or surfaces in dosimetry studies) with experimentally determined angular and energy distributions. This strategy is highly inefficient because a significant number of particles never intersect the detector. This paper proposes a refined Monte Carlo method to generate a sample of events that intersect the target volume, ensuring their angular distribution matches that of the conventional approach. It is based on the projection of a sphere containing the target volume onto a plane tangent to it at a fixed angle; this is termed the \textit{Probability Distribution Projection} (PDP) method. This configuration allows computation of the probability that a cosmic particle hits the sphere at this incoming angle, with this probability being proportional to the area of the corresponding section of a cylinder.
    The performance of this method demonstrates enhanced computational speed while yielding identical physical results. It has been implemented within the REST-for-Physics framework and tested using the geometry of a real detector, the IAXO-D0 Micromegas X-ray detector for the future axion helioscope BabyIAXO. The proposed method achieves a 37-fold improvement in efficiency compared to the traditional Monte Carlo scheme for the same accuracy, and is particularly advantageous when the target volume deviates from a spherical shape.
\end{abstract}


\maketitle

\section{Introduction} 

Cosmic ray simulations are used in a wide range of areas within particle physics. From gamma-ray astronomy to radiophysics, the interaction of these particles with matter must be modelled. Cosmic rays are composite showers of muons, photons, electrons, positrons, neutrons, protons, and light nuclei that are formed when the incident solar wind (primary protons from the Sun, but also other particles) interacts with the atmosphere. These constitute background components for any particle detector, with their contribution varying significantly depending on the detector's placement, materials, geometry and purpose.  

To quantify the background contribution of cosmic rays, numerical simulations are performed,  accounting for the detailed characteristics of the system under study. Generic particle detectors such as germanium gamma spectrometers \cite{hung2017investigation} or gaseous detectors for multiple purposes \cite{hohlmann2009geant4} are affected by this source of background. In rare event searches, such as interaction of dark matter particles or rare nuclear decays, cosmic-induced events constitute the major component of background \cite{aprile_conceptual_2014}; consequently, underground laboratories are crucial for these experiments. Neutrino telescopes also deal with this type of background \cite{carminati_atmospheric_2008}. Cosmic ray simulations are performed in dosimetry, for example to study their effect in high altitude flights where cosmic ray doses are computed for passengers, flight attendants and pilots \cite{yang2023simulation}. Additionally, the novel technique to probe large structures like volcanoes and pyramids, muongraphy \cite{nishiyama2016monte}, makes extensive use of cosmic ray simulations. Even the biggest particle physics experiments in particle colliders, like CMS or ATLAS, simulate their cosmic ray background \cite{sonnenschein_cms_2011, the_atlas_collaboration_studies_2011, boonekamp_cosmic_nodate}.

Many of these simulations follow a similar scheme: they generate the distributions of energy and direction of incident cosmic ray particles at a certain position on the Earth and altitude, then use these parameters to track the incident particles through matter, considering all possible interactions with their surroundings. 

Bridging these two steps presents a challenge, specifically, how to sample the incident particles for tracking. Typically, particles are sampled in a plane above the detector, adhering to the secondary distributions provided by cosmic ray generator software. Therefore, technically secondary particles produced by the interaction of cosmic rays with the atmosphere are simulated. The plane must be larger than the detector's geometry to allow for higher incident angles. However, this implies that a considerable number of sampled particles never interact with the detector because their trajectories do not intersect it. This is a well-known inefficiency in particle simulations that can be addressed by increasing the computation time.

In this work, we address this problem by introducing a geometrical scheme to transform the distributions of secondary particles from cosmic rays generated from a plane to the distributions of those that intersect with the detector. Consequently, all sampled particles interact with the objective volume, thereby increasing simulation efficiency and reducing computational costs. 

\subsection{Software for cosmic ray simulations}

Two distinct software programs are used for cosmic ray simulations: generators of cosmic ray particle distributions and detector simulation codes.

Examples of the first type of software packages are CORSIKA \cite{heck1998corsika}, CRY \cite{hagmann2007monte}, EXPACS \cite{sato2008development}, or MUTE \cite{woodley2024cosmic}. CORSIKA is a Monte Carlo simulation software specialized in air showers. It is capable of simulating particle secondaries from incident cosmic rays in the atmosphere. EXPACS and CRY differ somewhat; they model cosmic ray secondaries particle abundances using analytical formulas derived from accurate simulations. This enables rapid computation of radiation, considering the Earth's location and time of year. EXPACS simulations were done with PHITS Monte Carlo software \cite{sato2024recent}, and CRY is based upon MCNPX simulations \cite{TechReport_2022_LANL_LA}. Both demonstrate good agreement with each other and with experimental measurements. MUTE is a specialized software to model the cosmic particle flux underground. It models the effect of the materials and shapes above underground laboratories in order to estimate the remaining contribution of cosmic rays at certain depth.

The second type of software codes are used to simulate the interaction of the incident particles through matter. Here, detailed descriptions of the detector and the environment are needed. The fundamental approach involves using the output from a preceding software code to simulate the detector's response based on particle type, energy, interactions with surrounding materials, and other relevant factors. The most popular package with this role is Geant4 \cite{agostinelli2003geant4}, developed and maintained at CERN, but FLUKA \cite{ferrari2005fluka} is also quite extended.

\section{The PDP method}

\begin{figure}[h]
    \centering
    \includegraphics[width=0.45\textwidth]{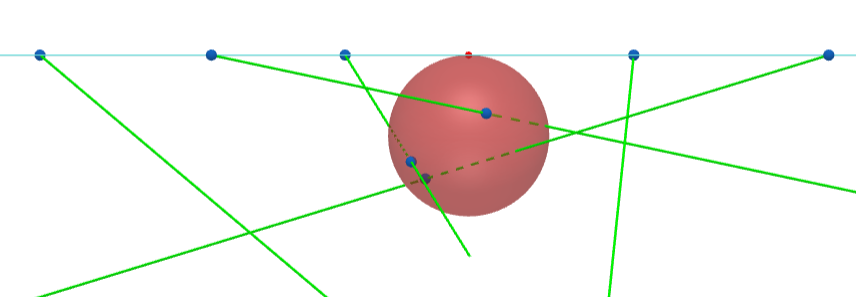}
    \caption{Geometrical configuration of the problem. Red sphere representing the detector, horizontal line is the tangent plane with primary positions in blue and particle trajectories in green.}
    \label{fig:PlaneSphere}
\end{figure}

The primary distinction in methodology between the Monte Carlo method and our proposal lies in the perspective from which cosmic ray parameters are described. In the Monte Carlo method, the simulation logic closely approximates the physical process. The particles considered in our detector simulation are the secondary cosmic rays resulting from the interaction of primary cosmic rays with the atmosphere. Since these secondary cosmic rays originate from above when the detector are at the surface, the simulation geometry is modelled with a horizontal plane on top of the detector from which the secondary particles are generated. The line that represents the trajectory of the particle is generated by choosing a random point in the plane and a random direction according to predefined probability density functions (figure \ref{fig:PlaneSphere}). If this line intersects the detector, represented here as a sphere, it is recorded; otherwise, it is discarded. This approach assumes that any secondary particle that does not intersect this virtual sphere enclosing our detector would not otherwise interact with the detector itself, thereby focusing computational effort on relevant trajectories. Consequently, numerous unnecessary trajectories, which do not intersect the sensitive volume, are simulated  when not using our optimization.

\begin{figure}[h]
    \centering
    \includegraphics[width=0.45\textwidth]{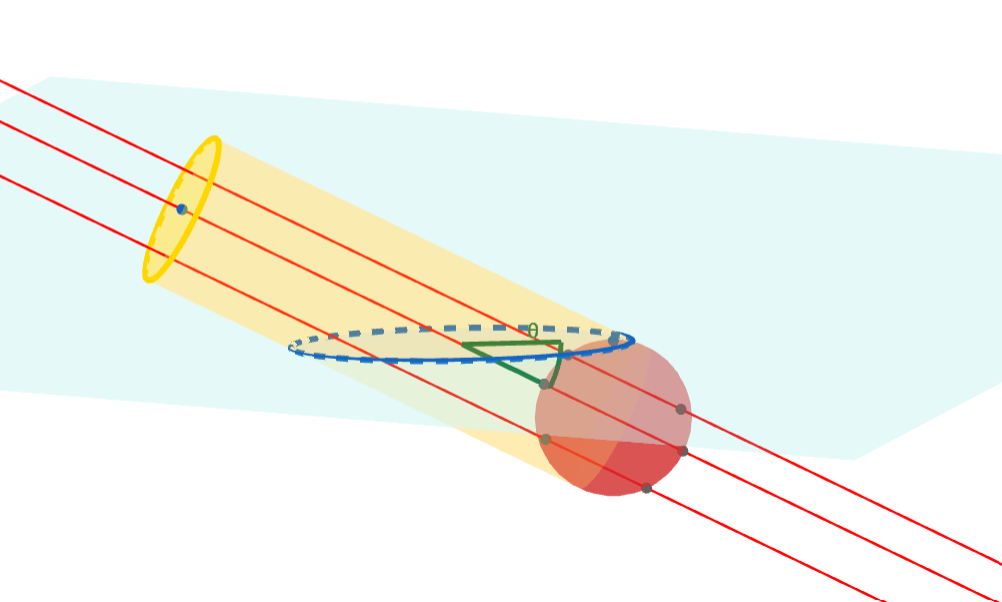}
    \caption{Grey plane tangent to the red sphere. In blue can be seen the projected ellipse from sphere for a given $\theta$, angle in green. All parallel rays, in red, inside the yellow cylinder have the same $\theta$ angle with respect to the plane and intersect the sphere.}
    \label{fig:PlaneSphere2}
\end{figure}

In our proposal, the logic starts from those cosmic rays that can be seen by the detector. Therefore, instead of simulating lines originating from the plane above, for each zenith angle, the origin of lines that intersect the sphere at this angle is considered. This implies that, for a fixed zenith angle $\theta$, the possible origin in the horizontal plane above the sphere is restricted to the area of the sphere's projection onto that plane at this specific $\theta$ angle, which invariably forms an ellipse (figure \ref{fig:PlaneSphere2}).
The area of the ellipse depends exclusively on the zenith angle, given fixed sphere dimensions, and can be readily computed as the area of a cylindrical section sliced at this angle $\theta$.

\noindent
\fbox{\parbox[b]{\linewidth}{

\textbf{Monte Carlo simulation steps:}

\begin{itemize}
    \item Sample position in plane above detector $\rightarrow$ $f(x,y)$
    \item Sample direction (zenithal and azimuthal angles) $\rightarrow$ $f(\theta, \phi)$
    \item Simulate trajectory.
\end{itemize}

Probability distribution of the direction, $f(\theta, \phi)$, is a multivariate distribution on two variables, zenithal and azimuthal angles, but very often both variables are independent. In that case, $f(\theta)$ and $f(\phi)$ can be considered.

Typically $f(x,y)$ and $f(\phi)$ are uniform distributions in their domains. However, $f(\theta)$ is not uniform and is commonly modelled as $f(\theta)=\cos^n(\theta)$, where $n \in \mathbb{Z}$. Additionally, empirical distributions for $f(\theta)$ can be extracted from measurements or simulations.

}}\\

This geometrical point of view is the key aspect because it also allows for the computation of the conditional probability density function for the azimuthal angle of lines that intersect with the sphere. If the projection area is larger, the probability of a line intersecting the sphere at this angle is also greater. This establishes a proportional relationship, meaning the probability is directly related to the area of the projected ellipse. Thus, the conditional probability density function can be constructed from the probability density function of zenithal angle in the plane as $f(\theta | \mathrm{Intersect\; sphere}) = f(\theta) \cdot (\mathrm{Area\; of\; projected\; ellipse,\; normalized})$.

While normalization of the probability distribution by the projected area would be desirable to obtain a valid probability density function, it is not strictly necessary for our purposes, as only the relative probabilities between angles are of concern. If 1000 particles are sampled using this method, only relative proportions between direction angles matter; therefore, the normalization process, which is non-trivial as the area of the projected ellipse is unbounded (e.g., infinite if $\theta=90^\circ$), is not critical. For practical applications, $\theta$ can be considered bounded, approaching but remaining below $90^\circ$, allowing the areas to increase significantly but never becoming infinite. The biggest area considered is used for normalization. Another advantageous aspect is that the incident distribution of zenith angles $f(\theta)$ experimentally approaches 0 as the angle nears $90^\circ$.

We named the method \textit{Probability Distribution Projection} (PDP) method because it involves the modification of the probability distribution in the zenith angle according to the projection of the sphere in the tangent plane.
\\

\noindent        
\fbox{\parbox[b]{\linewidth}{
\textbf{PDP method steps:}

\begin{itemize}
    \item Create distribution for zenith angle conditioned to rays that intersect the sphere: $f(\theta\; |\; \mathrm{Intersect\; sphere}) = f(\theta) \cdot (\mathrm{Area\; of\; projected\; ellipse})$
    \item Sample $f(\theta\; |\; \mathrm{Intersect\; sphere})$
    \item Sample azimuthal angle $f(\phi)$
    \item Sample a point in the projected ellipse.
    \item Compute the intersection point of the ray with the sphere.
    \item Simulate the event originating from the point on the sphere with the sampled direction.
\end{itemize}

}}\\

The conversion to equivalent simulated time is not straightforward for this PDP method cosmic generator. In Monte Carlo simulations, given the expected flux of incoming cosmic rays,  as all particles must be simulated, the equivalent time can be readily determined by dividing the total number of simulated particles (including those that hit the detector and those that do not) by the flux, for a fixed area. 

In the PDP method, the time required to produce the simulated number of particles must be re-scaled. Therefore, in order to obtain the equivalent time of the simulation, the number of simulated events for each zenith angle must be divided by $\cos(\theta)$. This represents the equivalent number of events that would have been simulated using the Monte Carlo method. To determine the time, the total area from the maximum allowed zenith angle has to be computed and the number of particles divided by the flux.

\section{Results}

\subsection{Implementation in Geant4}

This PDP method has been implemented within the Geant4 \cite{agostinelli2003geant4} simulation software through the interface framework REST-for-Physics \cite{altenmuller2022rest}. The algorithm operates within Geant4's \texttt{G4VUser PrimaryGeneratorAction::GeneratePrimaries} and utilizes a sampler to generate random values from an arbitrary probability distribution function, specifically a multivariate distribution function $f(E, \theta, \phi)$ in this instance. Typically, the azimuth angle $\phi$ can be considered independent and sampled separately.

In our implementation, which assumes azimuthal symmetry, the sampling of the multivariate distribution $f(E, \theta)$ is done via \texttt{TH2D::GetRandom2} for numerical distributions or via \texttt{TF2::GetRandom2} for analytical distributions. ROOT~\cite{brun1997root} is utilized due to its ready availability in REST-for-Physics, although any suitable sampling method may be employed. The ROOT random engine is initialized from Geant4's random engine to ensure reproducible results. Geant4's random engine is used for the remaining random numbers, such as the azimuth angle.

This probability density function from which samples are generated is the modified version derived from the previously described analytical projection. The zenith distribution for points in the horizontal plane is obtained from CRY and subsequently multiplied by the area of the projected ellipse for each $\theta$. 

The area of a section of a cylinder can be expressed as a function of the angle as \cite{harris1998handbook}:
$$
A = \pi r^2 \sec(\theta)
$$

For a unit sphere $r=1$, the constant $\pi$ can be discarded since the probability function will be normalized and only relative comparisons between angles matter.

Thus, the conditional probability function for the zenith angle of lines that intersect with the sphere is 

$$
f(\;\theta\; |\; \mathrm{Intersect\; sphere}) = f(\theta) \cdot \sec(\theta)
$$

\begin{figure}[h]
    \centering
    \includegraphics[width=0.45\textwidth]{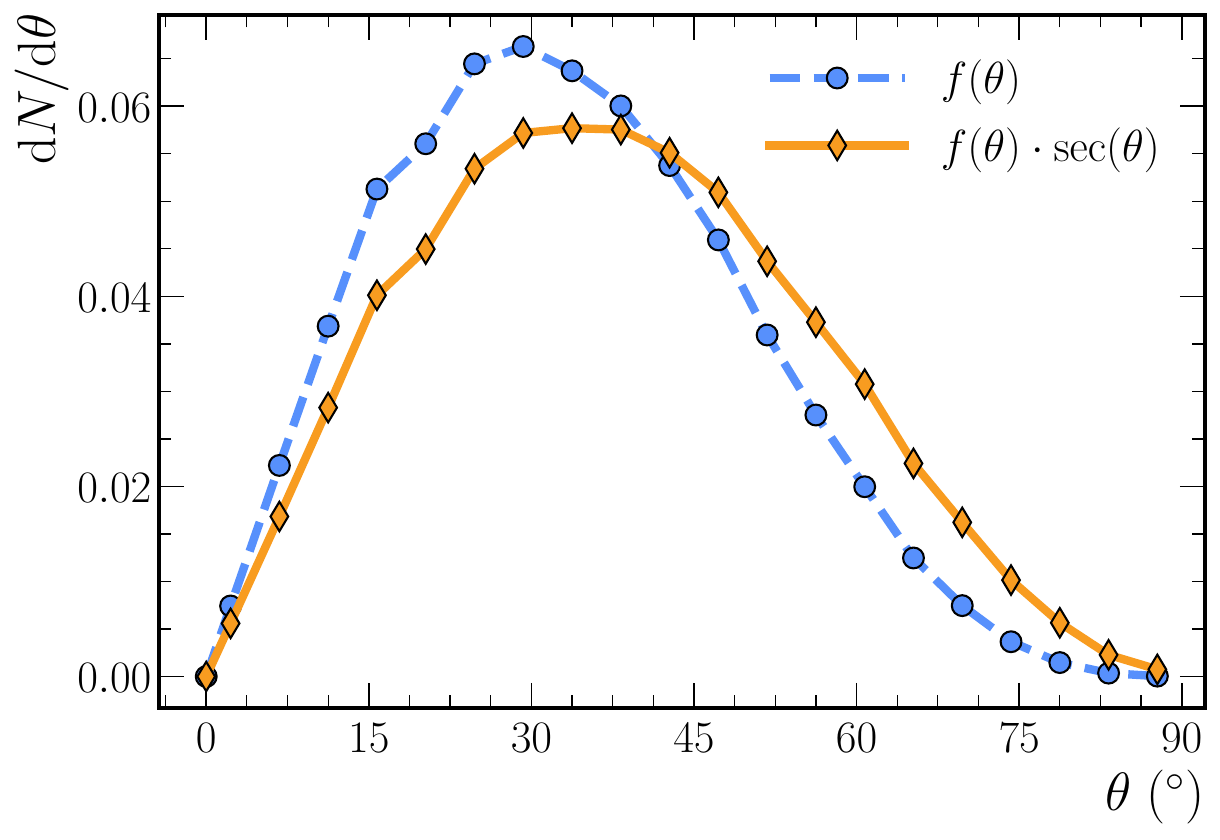}
    \caption{Probability density function $f(\theta)$ from CRY in blue (dashed) for muons at sea level (originated from proton primaries at latitude 41.6517º, sea-level altitude and date 1-1-2024) and modified conditional one, $f(\;\theta\; |\; \mathrm{ Intersect\; sphere})$ in orange (continuous).}
    \label{fig:PDF}
\end{figure}

This modified function also depends on energy; thus, $f(\theta)$ is more accurately represented as $f(E,\; \theta)$. Therefore, samples can be extracted from the 2-dimensional modified distribution $$f(E, \;\theta\; |\; \mathrm{Intersect \; sphere}) = f(E, \; \theta) \cdot sec(\theta)$$ see figure \ref{fig:PDF}. The distribution of $\phi$ is assumed uniform and independent from the other two variables, thus it is sampled independently.

After sampling the energy and direction, the intersection point between the line and the sphere is computed. This is a technical detail beneficial for Geant4 simulation, as all primaries will be approximately equidistant from the detector, in the surface of the sphere, but outside the boundaries of the simulated 'world', which is enclosed by the sphere.

\subsection{Performance of the PDP method}

\begin{table*}[t]
    \centering
    \resizebox{\linewidth}{!} {
    \begin{tabular}{lcccccc}
    \rowcolor[HTML]{356854} 
    {\color{white} Simulation} & {\color{white} Entries} & {\color{white} Primaries} & {\color{white} Saved (\%)} & {\color{white} Time (s)} & {\color{white} Yield (ent/s)} & {\color{white} Rate (ent/s)} \\ \hline
    \textbf{MC - R/r = 2.5} & 5M & 25.63M & 19.51 & 2635 & $1897 \pm 0.8$ & $0.529 \pm 0.0002$ \\
    \textbf{MC - R/r = 5} & 1M & 19.08M & 5.24 & 912 & $1097 \pm 1.1$ & $0.569 \pm 0.0006$ \\
    \textbf{MC - R/r = 10} & 1M & 75.30M & 1.33 & 2399 & $417 \pm 0.4$ & $0.576 \pm 0.0006$ \\
    \textbf{PDP method} & 1M & 1.00M & 99.997 & 395 & $2529 \pm 2.5$ & $0.577 \pm 0.0006$ \\
    \end{tabular}
    }
    \caption{Results for different simulation strategies in the simplest geometry: a sphere. Muons are sent from a plane/disk above the spherical sensible volume. This is the most favourable configuration possible for the PDP method, in more realistic simulations the improvement will be smaller. Monte Carlo simulations make use of a disk of radius $R$ from which muons are generated, $r$ denotes radius of the sphere containing the detector. 
    }
    \label{SphereTable}
\end{table*}

\begin{table*}[t]
    \centering
    \resizebox{\linewidth}{!} {
    \begin{tabular}{lcccccc}
    \rowcolor[HTML]{356854} 
    {\color{white} Simulation} & {\color{white} Entries} & {\color{white} Primaries} & {\color{white} Saved (\%)} & {\color{white} Time (s)} & {\color{white} Yield (ent/s)} & {\color{white} Rate (ent/s)} \\ \hline
    \textbf{MC - R/r = 2.5} & 1M & 85.06M & 1.18 & 3303 & $303 \pm 0.3$ & $0.383 \pm 0.0004$ \\
    \textbf{MC - R/r = 5} & 1M & 321.79M & 0.31 & 9864 & $101 \pm 0.1$ & $0.405 \pm 0.0004$ \\
    \textbf{MC - R/r = 10} & 0.36M & 460.26M & 0.08 & 14402 & $25.00 \pm 0.04$ & $0.407 \pm 0.0007$ \\
    \textbf{PDP method} & 1M & 16.93M & 5.91 & 1069 & $935 \pm 0.9$ & $0.409 \pm 0.0004$ \\
    \end{tabular}
    }
    \caption{As in Table \ref{SphereTable}, results for different simulation strategies in a realistic geometry: IAXO-D0 detector. This geometry is used to test the performance of the PDP method cosmic generator in a complex geometry against the traditional Monte Carlo method with different aspect ratios.}
    \label{IaxoTable}
\end{table*}

In order to show that the PDP method described in this work outperforms traditional Monte Carlo methods for cosmic ray simulations, several measurements in different geometries have been considered.

\begin{figure}[h]
    \centering
    \includegraphics[width=0.45\textwidth]{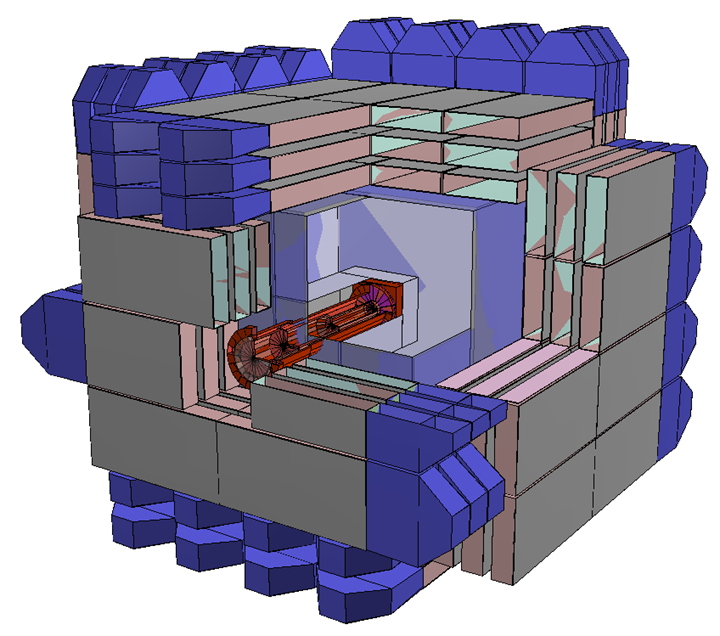}
    \caption{IAXO-D0 Micromegas X-ray detector geometry. In red the copper pipeline for X-rays and the TPC detector in the centre. The detector is surrounded by a veto system to tag muons and neutrons, consisting of three layers of plastic scintillators, in grey, with optic waveguides, in blue.}
    \label{fig:IAXO-D0}
\end{figure}

The PDP method has been developed with the abstract geometry of a sphere, hence this is the starting point to test the performance of the new method. The second should be a more complex geometry, a real detector: the IAXO-D0 Micromegas X-ray detector for the future axion helioscope BabyIAXO \cite{altenmuller2024background} (figure \ref{fig:IAXO-D0}). 
All muon background simulations shown in this work have been performed in GEANT4 using a single thread in the same computer (CPU Intel Core i9 13900K).

\begin{figure}[h]
    \centering
    \includegraphics[width=0.5\textwidth]{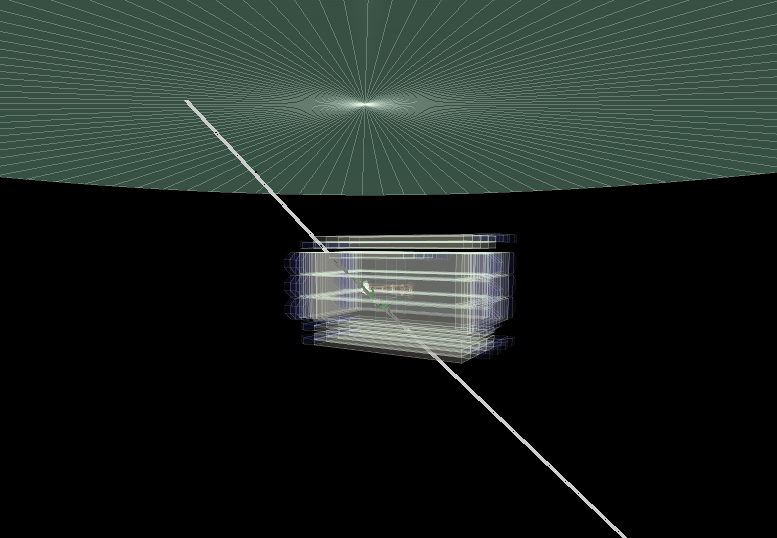}
    \caption{Monte Carlo simulation in Geant4. Particles are generated from the disk above the IAXO-D0 detector geometry.}
    \label{fig:IAXO_disk}
\end{figure}

Both geometries follow the same schema: cosmic rays are generated from a horizontal plane above the detector, this can be seen in figures \ref{fig:PlaneSphere}, \ref{fig:PlaneSphere2}, \ref{fig:IAXO_disk}. For practical purposes, this horizontal plane is modelled as a disk of radius $R$. The detector is encapsulated in a sphere of radius $r$. In the spherical detector used in the following simulations the radius is $r=5$ cm, and the IAXO-D0 detector is enclosed in a sphere of $r=180$ cm.

\begin{figure}[h]
    \centering
    \includegraphics[width=0.45\textwidth]{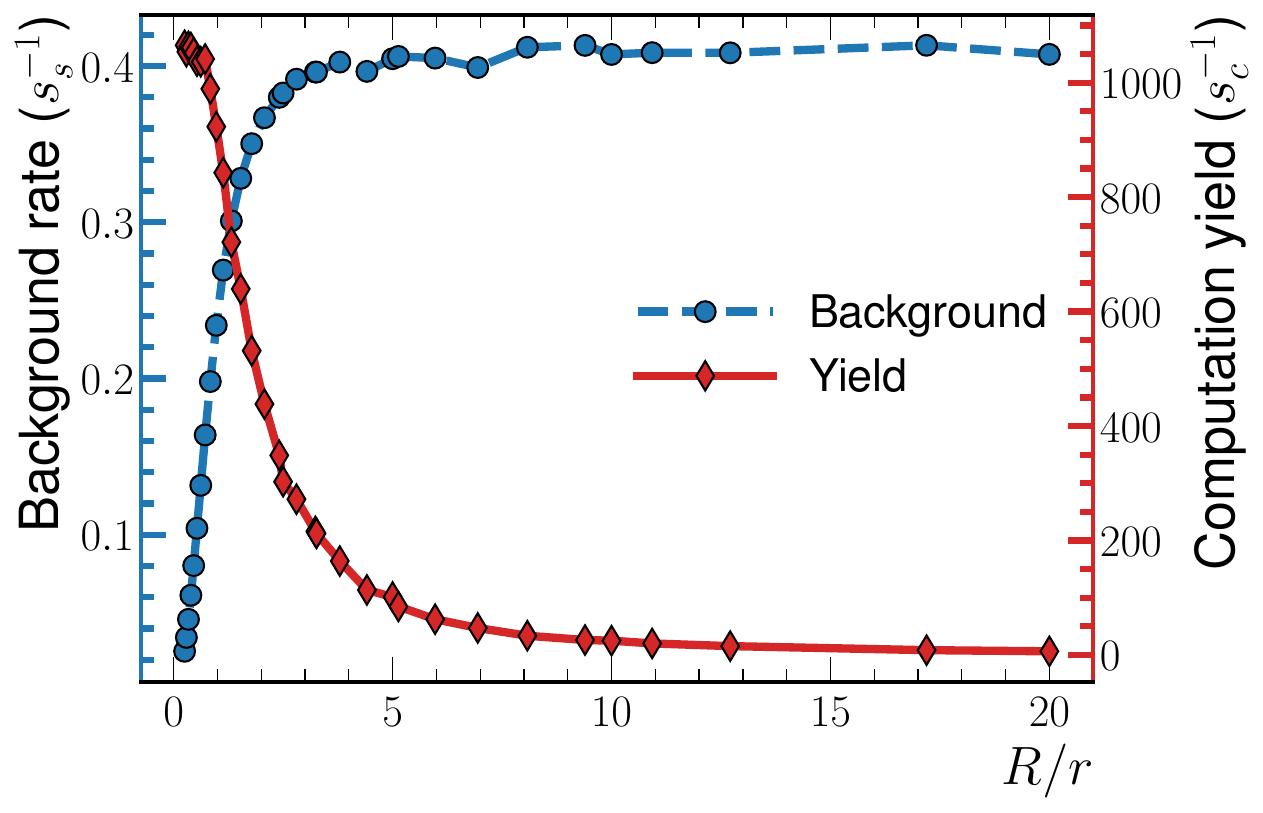}
    \caption{IAXO-D0 Monte Carlo simulation performances for different aspect ratio between disk radius, $R$, and sphere radius, $r$. For each geometry, the simulation is run until 10000 events are saved. The computation time, $s_c$, allows to compute the computation yield, the number of particles saved per second in each configuration, in red diamonds in the plot. Simulation time, $s_s$, is obtained from the area of the disk and the muon rate at surface. Here, the value 0.0055 $\mu ^-/s/cm^2$ has been used. In blue dots in the plot is shown this simulated background rate. In Monte Carlo simulations, a trade off is needed in order to select the most efficient aspect ratio for the accuracy needed. In this case, between 3 and 10 is reasonable.
    }
    \label{fig:disk-radius-study}
\end{figure}

The baseline method is a Monte Carlo random generating scheme. For this method,  the ratio $R/r$ has to be fixed. The relative size between the disk and the sphere is a balance between accuracy and computation speed. The larger the disk, the more accurate the simulation but the computation yield decreases as more particles have to be sampled to keep the same amount of events touching the detector. Using the IAXO-D0 geometry, 30 Monte Carlo simulations have been performed, varying the aspect ratio R/r from 0.25 to 20 (figure \ref{fig:disk-radius-study}). It can be seen that from $R/r = 5$, the level of accuracy reaches the plateau, meaning that the full momentum space is sampled (particles are allowed from higher angles). The number of particles per simulation time, $s_s$, is plotted in blue, the simulated background.
Increasing the aspect ratio only reduces the computation yield ($s_c$), which is defined as the saved particles per second and shown in red in figure~\ref{fig:disk-radius-study}.

From these Monte Carlo simulations, three promising values are identified, $R/r= $ 2.5, 5 and 10, and longer simulations are performed with both geometries, spherical and IAXO-D0 detectors, saving between 35 and 500 times more particles than before, from 10000 events to 350000, 1000000 or 5000000.
The results of these three longer Monte Carlo simulations can be seen in tables \ref{SphereTable} and \ref{IaxoTable}, for sphere and IAXO-D0 geometries respectively.

We now move to the PDP method. The same measurements are extracted from these simulations for both geometries. The PDP method developed in this work performs better in all aspects: best background rate, computation yield, percentage of saved events and computation time. The background rate is the physical particle rate that is being simulated, the number of particles saved per simulation time. Computation yield addresses the efficiency of the simulation, meaning the number of saved events per computation time, the real time that took the simulation.

Tables \ref{SphereTable} and \ref{IaxoTable} show this clearly, that the PDP method outperforms the Monte Carlo method in both geometries,it is faster and more efficient, even when compared with the fastest and least accurate Monte Carlo simulation with $R/r = 2.5$.

It is worth remarking that this method was derived with the purpose of simulating only the events that reach the detector. This can be seen in the spherical detector ($\%$ saved events), where almost all generated events are saved. In the case of IAXO-D0 detector, the abstract sphere with which the method was derived and through which all muons pass through is bigger than the sensitive volume and only events depositing energy in this volume are saved, resulting in different numbers of primaries and saved entries.

\subsection{Accuracy of the PDP method}

\begin{figure}[h]
    \centering
    \includegraphics[width=0.45\textwidth]{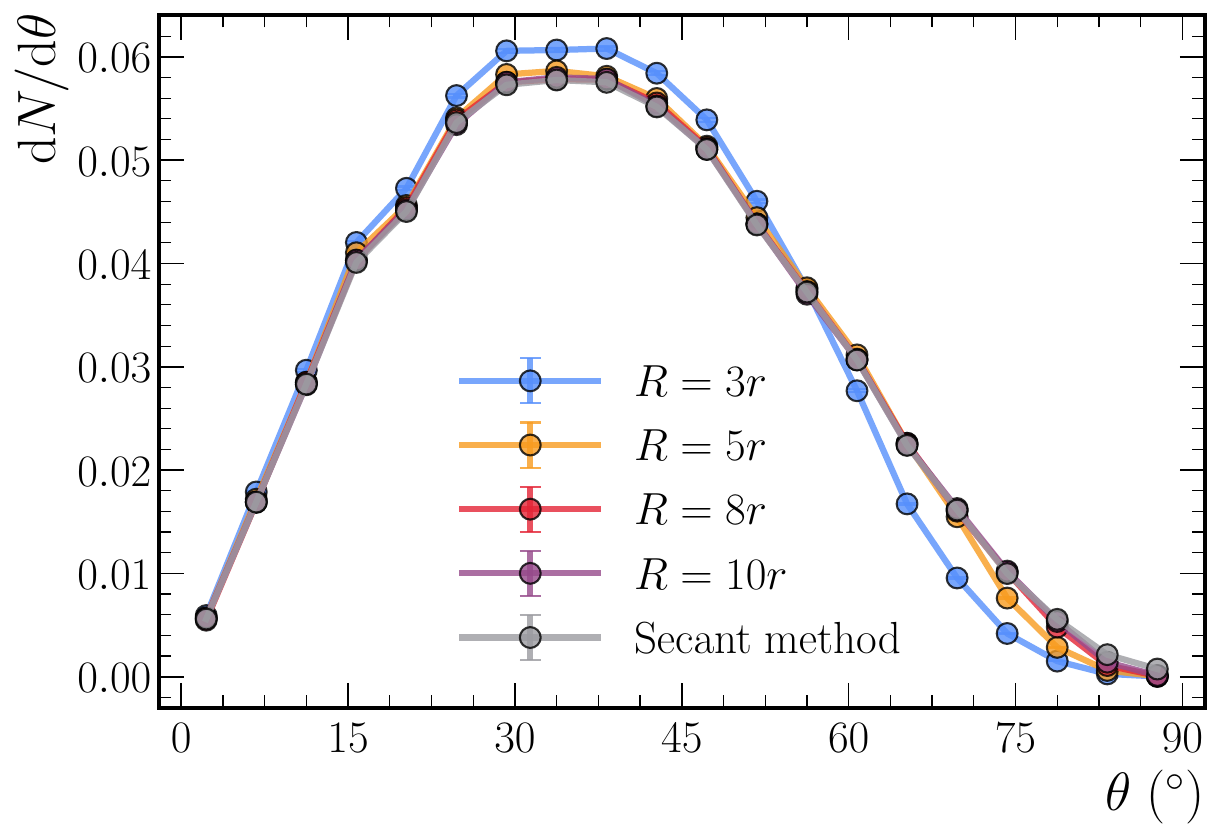}
    \caption{Zenith angle distribution for both methods. $R$ denotes the radius of the disk and $r$ the radius of the sphere. For this simulations, $r=180$ cm has been used, but only the relative value $R/r$ matters. High number of events makes statistical errors very small so they can not be appreciated in the plot.}
    \label{fig:AngleComparison}
\end{figure}

\begin{figure}[h]
    \centering
    \includegraphics[width=0.45\textwidth]{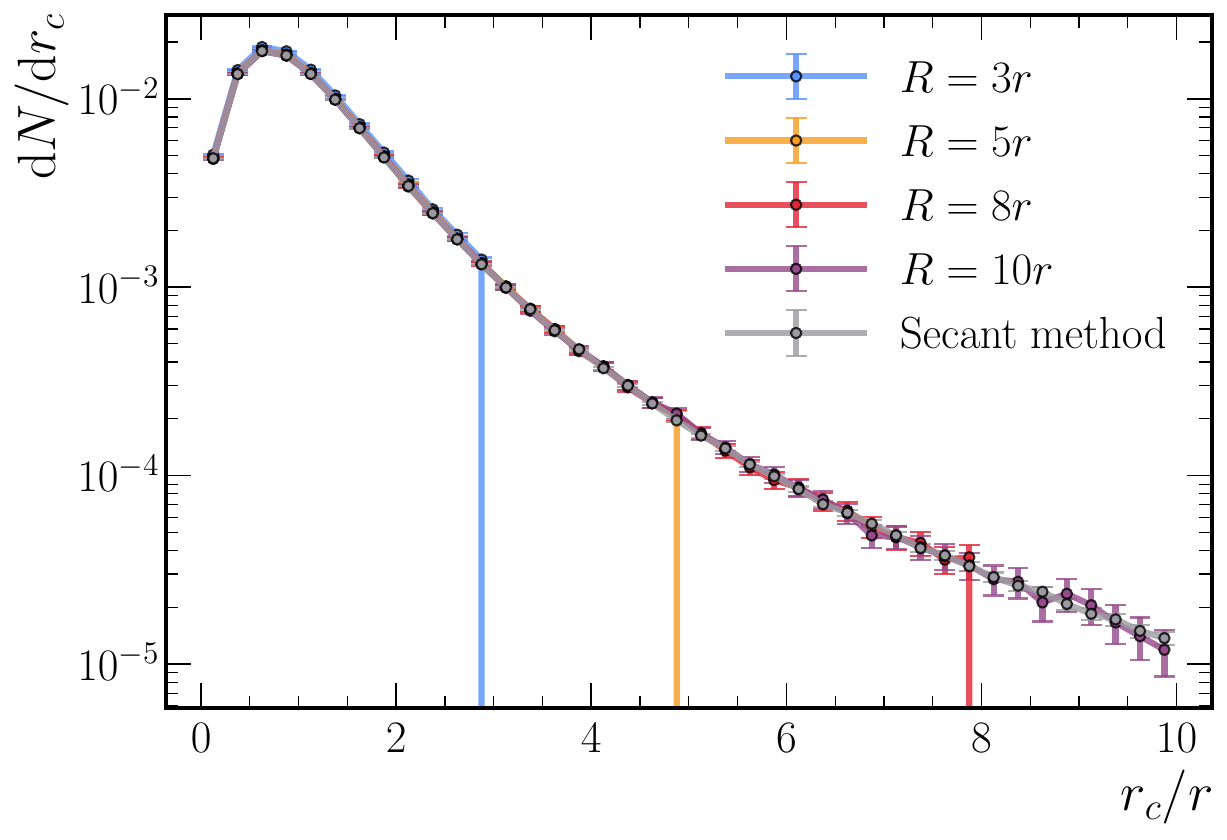}
    \caption{Distribution of the distance between the intersection point of the cosmic ray with the horizontal plane to the central axis, $r_c$, for both methods. It is expressed in number of sphere radii.}
    \label{fig:DistanceComparison}
\end{figure}

To check the fidelity of the PDP method two comparisons with the Monte Carlo method have been performed employing the more complex geometry (IAXO-D0).  First, the distribution for the conditional probability of zenith angle for rays that intersect with the sphere has been generated for both methods, see figure \ref{fig:AngleComparison}. The distribution for the Monte Carlo method approaches to the one of the PDP method cosmic generator as the radius of the sampling disc increases. The second cross-check is the distribution of the distance between the intersection point of the cosmic ray with the horizontal plane and the vertical axis that goes through the centre of the sphere, named distance-to-centre $r_c$. Measuring this parameter allows to compare the distribution of cosmic rays that produces this zenith angular distribution for both methods. As can be seen in figure \ref{fig:DistanceComparison}, the distribution of the distance $r_c$ is identical for both methods. In order to be able to compare different disk radii, all measurements are expressed in units relative to the radius of the sphere containing the geometry, $r$.

\section{Conclusions}

Simulations of the cosmic background are needed for the design of many types of particle detectors. For complex geometries, these simulations tend to be computationally expensive when using traditional strategies like the Monte Carlo method. 
In this work, we have developed a new method, called the PDP method, which optimises the computation through the simulation of only the particles that interact with the volume of interest, while keeping the statistical distributions of the Monte Carlo method. In some way, it is the translation to analytical expressions of the conditional probability from a Monte Carlo random schema.

Both methods have been tested using two geometries: a spherical volume and IAXO-D0 Micromegas X-ray detector in Geant4.
In the simplest configuration of a sphere, the PDP method saves all events (small discrepancy, 29 events in 1000000, due to the polygonal construction of the sphere in GEANT4), far better than any of the three Monte Carlo configurations, where the best value achieved is below 20$\%$ of saved events. Considering the geometry of a real detector like IAXO-D0, this efficiency decreases significantly (5$\%$ of saved events) because the detector is not spherical, however, it is still better than Monte Carlo (around 1 $\%$ at best). 

This increase in efficiency saving events is measured with the parameter \textit{computation yield}. It is the number of saved events per unit of computation time ($s_c$). This value is higher for the PDP method in both geometries: 1.33 times higher in the spherical geometry and 3 times higher in the IAXO-D0 geometry. 

Up to now, for these comparisons, the best values from all three Monte Carlo simulations are achieved for $R/r= 2.5$. This the fastest configuration in terms of computation yield. Still, accuracy should also be taken into account, as measured with the \textit{background rate}. It is the physical background rate that a real detector would see for a fixed muon flux. To measure this, the simulation time ($s_s$) has to be computed. In Monte Carlo simulations, it is derived from the area of disk. In the PDP method, it is computed by dividing the particle angle distribution, $f(\theta)$ by the flux times the area of the projected ellipse for each $\theta$ in the distribution. Simulated time is obtained integrating the resulting distribution. 

As shown in figure \ref{fig:disk-radius-study}, if the computation yield increases, the background rate decreases in the Monte Carlo method, so the simulation is less accurate, and some particles are missing. So, in these terms, the PDP method is also better than any of the Monte Carlo simulations. The background rate tends to a certain value for each geometry. In Monte Carlo simulations this is achieved by increasing the size of the disk, it can be observed in the green line in \ref{fig:disk-radius-study}. The PDP method always reaches this value, with both geometries tested close to and slightly above the best Monte Carlo background rate. 
So, taking into account accuracy, the proper comparison in terms of the yield of the calculation should also be done with $R/r=$10. In that case, PDP method is even better, being 6 and 37 times higher for spherical and IAXO-D0 detectors respectively.

Therefore, the PDP method developed in this work is faster (higher computation yield) and even more accurate (always achieve the proper background rate). 

\section*{Acknowledgements}

We acknowledge the support of the European Union's Horizon 2020 Program, under the European Research Council (ERC), grant agreement ERC-2017-AdG788781 (IAXOplus) and from 'European Union NextGenerationEU / PRTR' (Planes complementarios, Programa de Astrofísica y Física de Altas Energías). Finally, we also acknowledge support from Gobierno de Aragón through their predoctoral research contracts, competitive call 2020-2024.


\section{}\label{}

\printcredits


\printbibliography



\end{document}